\documentclass[preprint,showpacs,amsmath,amssymb]{revtex4}
\usepackage{amsmath,amssymb}
\usepackage{graphicx}

\begin{document}

\title{Long-range strain correlations in sheared colloidal glasses}
\author{Vijayakumar Chikkadi$^1$, Gerard Wegdam$^1$, Daniel Bonn $^1$, Bernard
Nienhuis$^2$, Peter Schall$^1$.}

\affiliation{
$^{1}$ Van der Waals-Zeeman Institute, University of Amsterdam, Science Park 904,
1098 XH Amsterdam, The Netherlands.\\ $^{2}$ Institute for Theoretical Physics,
University of Amsterdam, Science Park 904, 1098 XH Amsterdam, The Netherlands.}

\begin{abstract}
Glasses behave as solids on experimental time scales due to their slow relaxation. Growing dynamic length scales due to cooperative motion of particles are believed to be central to this slow response. For quiescent glasses, however, the size of the cooperatively rearranging regions has never been observed to exceed a few particle diameters, and the observation of long-range correlations that are signatures of an elastic solid has remained elusive. Here, we provide direct experimental evidence of long-range correlations during the deformation of a dense colloidal glass. By imposing an external stress, we force structural rearrangements that make the glass flow, and we identify long-range correlations in the fluctuations of microscopic strain, and elucidate their scaling and spatial symmetry. The applied shear induces a transition from homogeneous to inhomogeneous flow at a critical shear rate, and we investigate the role of strain correlations in this transition.
\end{abstract}


\maketitle

Glasses have attracted considerable attention due to their ubiquity in nature, and
for the scientific challenges they pose \cite{glasses}. Due to their long
relaxation time, glasses behave as solids on experimental time scales; this slow
response is often attributed to cooperative motion of particles
\cite{adam_gibbs_65}. Dynamic correlations that are used to characterize the
length scales of the cooperatively rearranging regions are indeed observed to grow
on approach to the glass transition
\cite{simulation_dc,colloids_expt}. However, despite the rapid growth of the
glass relaxation time and the glass viscosity by many orders of magnitude, the
length scale associated with the dynamic correlations increases only weakly on
approach to the glass transition. Simulations as well as experiments have shown
that in quiescent glasses, the dynamic length scales remain limited to a few
particle diameters
\cite{simulation_dc,colloids_expt}. On the other hand, recent simulations on
driven, athermal amorphous materials have evidenced the existence of
avalanche-like, long-range correlated flow \cite{Correlations_Athermal}.
Correlations in slowly driven thermal glasses, however, are poorly explored. Here,
we show that even a glass exhibits long-range correlations in the microscopic flow
field when it is driven so slowly that the rearrangements induced by the flow
happen on a timescale similar to that of the spontaneous (thermal) rearrangements.
By imposing very slow shear, we explore the transition between a regime where the
spontaneous fluctuations are sufficiently rapid to accommodate the flow, to a
regime where this is no longer possible, and this allows us to probe long-range
correlations. Nevertheless, in strained molecular glasses, direct observation of
such correlations is prohibitively difficult, because the small size of molecules
inhibits direct observation. We therefore use direct real-space observation in
suspensions of colloidal particles; the individual particles can be imaged and
their motion be tracked accurately in three dimensions with confocal microscopy
\cite{weeks_weitz_00}. Hard-sphere colloidal suspensions have become a much-used
model system for studying glasses; at high packing densities, the motion of the
colloidal particles becomes increasingly frustrated, and structural relaxations
slow down dramatically at particle volume fractions larger than $\phi_g \sim
0.58$, the colloidal glass transition \cite{pusey_megen}. By taking advantage of
recent imaging techniques to image large sample volumes containing as many as
$\sim 2.5
\times 10^5$ particles, we show that the flow of glasses is governed by
surprisingly long-range correlations .

We follow correlations in the deformation of the glass directly in three
dimensions and real time by visualizing fluctuations in the microscopic strain and
non-affine displacements. These strain fluctuations exhibit remarkably long-range
correlation that extend over the full system size, far beyond the range of dynamic
correlations reported for quiescent glasses.
\cite{simulation_dc,colloids_expt}. We investigate these correlations during
two different modes of macroscopic deformation: during homogeneous flow at very
low applied shear rates $\dot{\gamma}$ much smaller than the inverse structural
relaxation time $\tau^{-1}$ of the glass, and during inhomogeneous flow at strain
rates $\dot{\gamma}>\tau^{-1}$. These two modes of deformation are also observed
for molecular glasses \cite{shear_bands}, and the direct visualization of strain
fluctuations allows us to identify their important role in the transition between
both modes. Surprisingly, we find that the scaling behavior of the strain
correlations is uniform over the investigated range of shear rates, suggesting a
robust, scale-free organization of the flow of glasses.

We used sterically stabilized fluorescent polymethylmethacrylate (PMMA) particles
suspended in a density and refractive index matching mixture of Cycloheptyl
Bromide and Cis-Decalin. The particles have a diameter of $\sigma = 1.3 \mu m$,
with a polydisperity of $ 7 \%$ to prevent crystallization. A colloidal glass with a volume fraction of $\phi \sim 0.60$ is prepared by diluting samples centrifuged
to a sediment with $\phi \sim 0.64$. The suspension is loaded in a cell between
two parallel plates a distance of $65 \mu m$ apart. Boundary-induced crystallization is suppressed by a layer of polydisperse particles on the plates. We used a piezoelectric translation stage to move the top plate to apply shear at very slow rates between $\dot{\gamma} = 10^{-5} s^{-1}$ and $\dot{\gamma} = 10^{-4} s^{-1}$, of the order of the inverse structural relaxation time of the glass. We used confocal microscopy to image individual particles in a $107 \mu m$ by $107 \mu m$ by $65\mu m$ volume, and determined their positions in three dimensions with an accuracy of $0.03\mu m$ in the horizontal, and $0.05\mu m$ in the vertical direction \cite{weeks_weitz_00}. We tracked the motion of $\sim 2 \times 10^5$ particles during a $25~min$ time interval by acquiring image stacks every $60~sec$. The mean square displacement $<\Delta r^2>$ of particles in the quiescent glass is shown in Fig. 1a; we estimate the structural relaxation time of the glass $\tau \sim 2\times 10^4$ sec from the requirement $<\Delta r(\tau)^2>=(\sigma/2)^2$; this relaxation time is a factor of $~5\times 10^4$ larger than the Brownian relaxation time at infinite dilution, $\tau_B=0.4s$. Because the hard-sphere glass exhibits aging, this structural relaxation time changes with time. To obtain reproducible results, all measurements presented here were taken consistently $\sim3$ hours after mechanical rejuvenation of the sample.

We study correlations in the deformation of the glass by decomposing the particle
motion into affine and non-affine components. To do so, we follow particle
trajectories and identify the nearest neighbors of each particle as those
separated by less than $r_{0}$, the first minimum of the pair correlation
function. We subsequently determine the best affine deformation tensor ${\bf
\Gamma}$ describing the transformation of the nearest neighbor vectors, ${\bf
d_i}$, over the time interval $\delta t$
\cite{falk_langer_98}, by minimizing $D^2_{min} = (1/n) {\sum_{i=1}^{n}}({\bf
d_i}(t + \delta t) - {\bf \Gamma}{\bf d_i}(t))^2$. The symmetric part of ${\bf
\Gamma}$ is the local strain tensor. The remaining non-affine component
$D^2_{min}$ has been used as a measure of plastic deformation
\cite{falk_langer_98}. We define correlations in the fluctuations of the
principal shear strain component $\epsilon_{xz}$, and the non-affine displacement
$D^2_{min}$ using:
\begin{equation}
C_A({\bf \delta r}) = \frac{ \left< A({\bf r + \delta r}) A({\bf r})
\right> - \left< A({\bf r}) \right> ^{2} } { \left< A({\bf r})^{2}
\right> - \left< A({\bf r}) \right> ^{2} }  ,
\label{c_r}
\end{equation}
where $A=\epsilon_{xz}$ or $A=D^{2}_{min}$, and angular brackets denote ensemble
averages. $C_A$ correlates values of $\epsilon_{xz}$ or $D^{2}_{min}$ at locations separated by ${\bf \delta r}$.

We investigate strain correlations during homogeneous flow of the glass by
subjecting the glass to a shear rate of $\dot{\gamma} \sim 1.5 \times 10^{-5}s^{-1}$, a factor of 6 smaller than the inverse structural relaxation time.
At this shear rate, thermally activated relaxation occurs sufficiently fast, and
after an initial transient the colloidal glass flows homogeneously. We shear the
glass to $\gamma \sim 1$ to address steady-state flow, and show the displacements
$\Delta x$ of the particles during a small shear increment as a function of height in Fig. 1b. The average particle displacement (dashed red line) increases linearly with height, indicating uniform flow. Strong fluctuations exist, however, at the
level of the individual particles. We determine values of the local shear strain
from the motion of a particle with respect to its nearest neighbors during time
intervals of $3~min$ and $25~min$ and plot the relative frequency of occurrence of shear strain values as a function of shear strain magnitude in Fig. 1c, inset. For short times, the probability distribution of strain values decreases as a
power-law over almost three orders of magnitude in probability. Such power-law
decay is a fingerprint of highly correlated dynamics \cite{crystals}. This is also evident from the real-space distribution of the microscopic shear strain
$\epsilon_{xz}$ (Fig. 1c). Red regions indicate zones where large shear strain is
localized and irreversible rearrangements occur, also known as shear
transformation zones
\cite{falk_langer_98,schall_spaepen_07,shear_bands}. We determine
correlations between these zones by calculating the correlation functions of the
shear strain, $C_{\epsilon_{xz}}$, and of the non-affine displacements,
$C_{D^{2}_{min}}$. At the short time intervals considered here, the correlation
functions do not depend significantly on the specific time interval chosen, and
appear robust. Correlations in the $x$-$z$ plane are obtained by taking ${\bf
\delta r} = (\delta x,0,\delta z)$; a corresponding color coded representation of
the correlation function $C_{\epsilon_{xz}}$ is shown in Fig. 1d. Remarkably, the
correlation function shows a four-fold pattern at its center, which reveals the
elastic response of the glass to local shear transformation zones
\cite{quad_strain}. This glass elasticity leads to strong correlations between
shear transformation zones as evidenced by the regular pattern of yellow zones.
This is most evident if we average $C_{\epsilon_{xz}}$ within angular wedges
around the horizontal, vertical, and diagonal directions, and plot the
corresponding angle-specific correlation function versus $r$ in Fig. 1e. The
correlation function $C_{D^2_{min}}$ appears to be isotropic. We average over all
directions and plot the magnitude of $C_{D^2_{min}}$ as a function of distance in
Fig. 1f. A remarkable power-law decay is observed, which is truncated at the
vertical system size, $\delta r / \sigma\sim 50$; thus the correlations span the
entire system. We further confirmed that these correlations are independent of the
specific measure of non-affinity by using different definitions of non-affine
motion. The correlation function remained robust. These results provide direct
evidence of the existence of long-range strain correlations in a glass, and they
highlight the scale-free character of the non-affine rearrangements that govern
plastic deformation. The scale invariance appears to be a generic feature of
elasto-plastic deformation in other materials, too: the dislocation motion in
crystals \cite{crystals}, and the aftershocks in earth quakes \cite{earthquake}
display similar scale-free patterns.

We probe these strain correlations by subjecting the glass to increasing shear
rates. While the flow remains homogeneous over a range of low strain rates, at a
critical strain rate of $\dot\gamma_c \sim \tau^{-1}$ we observe a sudden
transition to inhomogeneous flow. The glass separates into two bands that flow at
different rates, as illustrated by the particle displacements as a function of
height obtained at a shear rate of $\dot{\gamma}\sim 1 \times 10^{-4}s^{-1}$ (Fig. 2a). Linear fits to the displacement profile for $z<23 \mu m$ and $z>27 \mu m$ (dashed lines in Fig. 2b) yield strain rates of $4 \times 10^{-5} s^{-1}$ and $2.2 \times 10^{-4} s^{-1}$, respectively, that differ by a factor of five. A reconstruction of the shear strain distribution shows that highly non-affine shear transformation zones accumulate in the upper part (Fig. 2b). We investigate the robustness of the scaling observed in Fig. 1e by determining $C_{D^2_{min}}\left( \delta r \right)$ separately for the high and low shear band. The resulting angle-averaged correlation functions are shown together with those of homogeneous flow in Fig. 2c. A remarkable collapse of the data is observed. While the magnitude of fluctuations in the two bands differs largely, the normalized correlation function shows very similar power law decay (Fig. 2c): the same scaling exponent applies to the low and the high shear band, as well as to homogeneous flow. We find a scaling exponent of $\alpha=1.3 \pm 0.1$
from the best fit to the data. Athermal and quasi-static shear simulations of
amorphous solids \cite{maloney_lemaitre} have shown similar long-range
correlations; however, the effect of finite shear rate and finite temperature on
the statistical correlations between shear transformation zones remained unclear
\cite{finite_shear_temp}. Our results conclusively show long-range correlations
even at finite shear rate, and at finite temperatures.

The difference between the two bands becomes evident when we investigate particle
dynamics as a function of time. The mean square displacement of particles around
the mean flow, $<{r'^2}(t)> = <({\bf \Delta r}(t) - \left<{\bf \Delta r}(t)\right>_z)^2>$, where $\left<{\bf \Delta r}(t)\right>_z$ is the average displacement at height $z$, reveals mobilities in the two bands (Fig. 2d) that are larger than those at rest (Fig. 1a), and that differ significantly between the two bands. For particles in the high shear band, the mean square displacement is significantly larger indicating enhanced particle diffusion compared to particles in the low shear band. Remarkably, the strain correlation function, $C_{\epsilon_{xz}}$, calculated separately for both bands (Figs. 2e and f) reveals a spatial symmetry change. While for the low shear band, the central four-fold symmetry is still predominant (Fig. 2e), for the high shear band, this symmetry is lost, and the pattern appears isotropic (Fig. 2f). This symmetry change reflects the transition from a solid to a liquid-like response of the glass. These results highlight the central importance of strain correlations: While the transition from a reversible elastic to an irreversible viscous response is usually associated with a symmetry change in time, our results show that this transition is also associated with a spatial symmetry change in the strain correlation function, underlining its fundamental character. A similar interpretation is given to fracture surfaces of metallic glasses that display striking evidence of such a solid to liquid transition \cite{spaepen_75}. Our colloidal glass allows us to directly visualize the strain correlations and identify their central importance in this solid-to-liquid transition.

How does this transition emerge? To elucidate this, we follow correlations during
the initial stages of shear banding. We determine strain distributions during the
transient stages, before shear bands manifest, and calculate correlation functions
for the entire field of view (Figs. 3a and b). In the early stage, the correlation
shows a strong bias in the horizontal direction (arrows). This horizontal bias
signals the excitation of additional elastic modes at higher shear rates that
cause strong correlations between shear transformation zones in the horizontal
direction. This bias lowers the effective resistance to flow in the direction of
shear, thereby leading to shear bands in the later stages of deformation (Fig.
3b). These results highlight the importance of long-range correlations in the
shear banding of glasses. Shear experiments at much higher shear rates show that shear bands can originate from density fluctuations \cite{besseling_2010}. The model proposed by the authors suggests a linear viscous flow at very low shear rates that is hard for them to address experimentally. Here, we have successfully addressed this low shear rate regime; our measurements at the onset of shear banding show strong changes in the correlations of microscopic strain, but no measurable dilatancy.

Our results establish the existence of long-range strain correlations in the flow
of glasses. The long-range elastic interactions between shear transformation zones
lead to scale-free deformation of the glass. While our results for shear banding
are obtained for colloidal glasses, they should be generic to glassy flows. The
formation of shear bands has often been linked to strain softening of the
material, caused by excess dilation, that accompanies the formation of shear
transformation zones \cite{shear_bands, besseling_2010}. The direct imaging of
strain correlations that we have reported here demonstrates that long-range
elastic correlations play a central role in the manifestation of such
instabilities
\cite{bulatov_argon_94}. Finally, the robust scaling that we observe suggests
a naturally scale-free flow and relaxation of glasses. We propose similar analysis
of shear flows in systems like granular, foams and emulsions to test the
universality of the scaling exponent, and to determine the universality class of
the flow and relaxation of amorphous materials.

We thank D. Frenkel, and F. Spaepen for helpful discussions. This work was
supported by the Innovational Research Incentives Scheme ("VIDI" grant) of the
Netherlands Organization for Scientific Research (NWO).

\newpage
\begin{figure}

\includegraphics[width=0.7\textwidth]{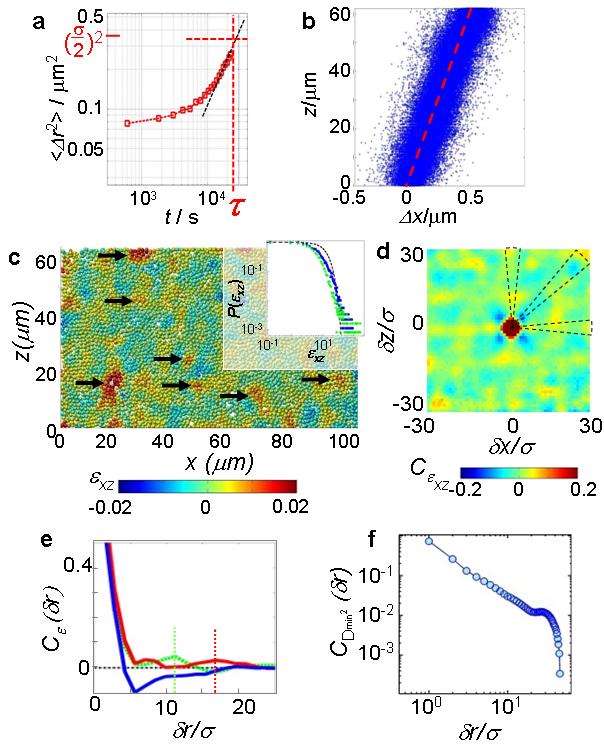}
\caption{Homogeneous deformation at a shear rate of $\dot{\gamma} = 1.5 \times
10^{-5} s^{-1}$.
(a) Mean square displacement of the quiescent glass without applied shear. The
structural relaxation time, $\tau$, of the glass is estimated by extrapolating the
mean square displacement in the diffusive regime with a line of slope one (dashed
black line). (b) Displacements of individual particles (+) and average
displacement (dashed line) along the shear direction during $\delta t=10 ~min$ of
shear. (c) $7 \mu m$ thick reconstruction of the distribution of shear strain
after $\delta t=3~min$ of shear. Particle color indicates the value of
$\epsilon_{xz}$. Inset in (c) shows the relative frequency of shear strain
magnitudes $\epsilon_{xz}$ for time intervals $\delta t=3~min$ (green stars) and
$25~min$ (blue squares). (d) Angle resolved spatial correlation,
$C_{\epsilon_{xz}}$, of the fluctuations of shear strain, in the $x$-$z$ plane.
(e) $C_{\epsilon_{xz}}$ as a function of distance averaged over angular wedges of
$10^{\circ}$ around the horizontal (red line), the vertical (green line), and the
two diagonal directions (blue line). (f) Angle-averaged correlation function
$C_{D^2_{min}}$ as a function of distance in a double-logarithmic
representation.}
\label{fig_1}
\end{figure}

\begin{figure}
\includegraphics[width=0.7\textwidth]{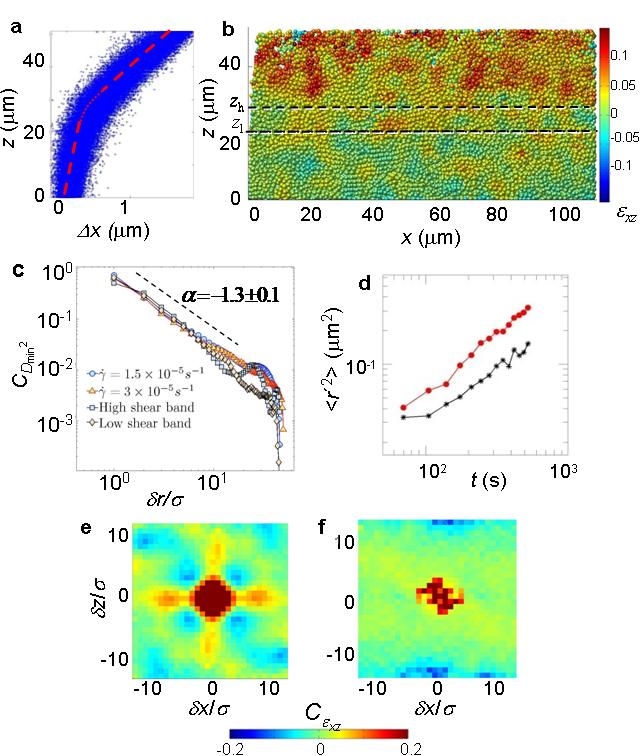}
\caption{Inhomogeneous deformation at a shear rate of $\dot{\gamma} = 1 \times
10^{-4} s^{-1}$.
(a) Particle displacements along the shear direction during $\delta t = 4 ~min$ of
shear. Dashed red lines are linear fits to the shear profiles for $z < z_l = 23
\mu m$ (low shear band) and $z > z_h = 28 \mu m $ (high shear band). (b) $7 \mu m$
thick reconstruction of the distribution of incremental shear strain
$\epsilon_{xz}$ during the time interval $\delta t=7 ~ min$. (c) Angle-averaged
correlation function $C_{D^2_{min}}$ as a function of distance $\delta r$, for the
low shear band (blue squares), the high shear band (yellow diamonds), and for
homogeneous shear at $\dot{\gamma}= 1.5 \times 10^{-5 } s^{-1}$ (blue dots) and $3
\times 10^{-5 } s^{-1}$ (orange triangles). A least square fit to the data gives a
slope of $\alpha \sim -1.3
\pm 0.1$ (dashed line). (d) Mean square displacement of particles in the low
shear band (black stars and line) and high shear band (red dots and line). (e and
f) Angle-resolved spatial correlations of shear strain, $C_{\epsilon_{xz}}(\delta
r)$, in the $x-z$ plane, for the low and the high shear bands, respectively.
Correlations are computed over a time interval of $\delta t = 3~min $.}
\label{fig_2}
\end{figure}

\begin{figure}
\includegraphics[width=0.7\textwidth]{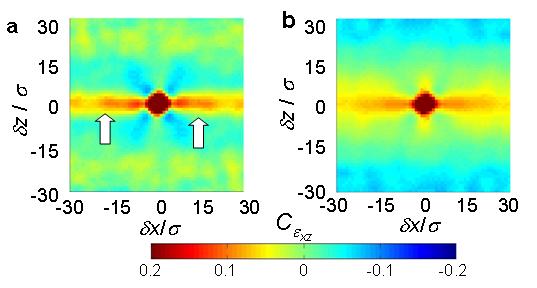}
\caption{Evolution of strain correlations during shear band formation.
Shear strain correlation functions $C_{\epsilon_{xz}}$ computed for the entire
field of view before (a) and after (b) the manifestation of shear bands. The
arrows in (a) indicate the strong correlation in the direction of shear that lead
to shear banding in later stages (b).}
\label{fig_3}
\end{figure}

\end{document}